\documentstyle[12pt,epsfig]{article}
\begin{document}
\thispagestyle{empty}
\setcounter{page}0

~\vfill

\begin{center} {\LARGE\bf New Tensor Particles \ \\
\ \\
from $\pi^- \rightarrow e^- \tilde{\nu} \gamma$ and
$K^+ \rightarrow \pi^o e^+ \nu$ Decays} \end{center}
\ \\
\ \\
M.~V.~Chizhov\\
{\it Centre of Space Research and Technologies, Faculty of Physics, University
of Sofia, \\
1126 Sofia, Bulgaria}

\vfill

\begin{abstract}
The inserting of antisymmetric tensor fields into the standard electroweak
theory may explain the recent experiments on $\pi^- \rightarrow e^- 
\tilde{\nu} \gamma$ and $K^+ \rightarrow \pi^o e^+ \nu$ decays. New
intermediate particles can induce the destructive interference in the pion
decay and the nonzero tensor and scalar form factors in the kaon decay.
\end{abstract}

\vfill

\newpage
\ \\
\ \\
\ \\
\ \\
\pagestyle{myheadings}
\markboth{empty}{}

\section{Introduction}

Recently the interest in possible tensor interactions in weak decays increased.
This is connected with the last experiments on$^1$  $\pi^- \rightarrow
e^- \tilde{\nu} \gamma$ and$^2$  $K^+ \rightarrow \pi^o e^+ \nu$ decays
(see also Ref.~3). The experimentally
obtained form factors cannot be explained in the framework of the standard
electroweak theory. This led to numerous discussions on the possible existence
of tensor terms in the effective Fermi interaction.$^{4-7}$ In Ref.~4
results of the experiment on $\pi^- \rightarrow e^- \tilde{\nu}
\gamma$ decay were explained by introducing an additional
tensor interaction in the Fermi Lagrangian
\begin{equation}
{\cal{L}}_T =  \sqrt 2 G_F V_{ud} f_T\;\bar{u}_R \sigma_{\mu\nu} d_L \cdot 
\bar{e}_R \sigma^{\mu\nu} \nu_L,
\end{equation}
where $V_{ud}$ is an element of the Kobayashi-Maskawa mixing matrix and
$\sigma_{\mu\nu} = {i \over 2}\; [\gamma_{\mu}, \gamma_{\nu}]$. The constant 
$f_T$ is estimated in the framework of the relativistic quark model:
$f_T=-(4.2\pm1.3)\times10^{-2}$.
The independent calculation$^5$ by QCD techniques leads to
\mbox{$f_T=-(1.4\pm0.4)\times10^{-2}$}. 
It was demonstrated in Ref.~6 that
using (1) an agreement between all data of the previous experiments on$^8$  
$\pi^- \rightarrow e^- \tilde{\nu} \gamma$ can be obtained. However,
in Ref.~7 a mechanism for generating a (pseudo)scalar interaction
\begin{equation}
{\cal{L}}_S = \sqrt 2 G_F V_{ud}f_0\;(\bar{u}_R d_L)(\bar{e}_R \nu_L)
\end{equation}
from (1) was presented leading to $\mid f_T \mid <10^{-4}$, when the
restrictions on $f_0$ from $\pi_{e_2}$ decay are accounted for $^9$:
$\mid f_0 \mid < 2.8 \times 10^{-6}$.

In this paper we demonstrate that the simultaneous description of the
results from the meson decays experiments is possible if 
the standard electroweak theory is extended by an additional Higgs doublet
and two doublets of antisymmetric tensor particles:
$(T^+_{\mu\nu} T^o_{\mu\nu})$ and $(U^o_{\mu\nu} U^-_{\mu\nu})$.
This allows to explain both the destructive interference in the amplitude of
the $\pi^- \rightarrow e^- \tilde{\nu} \gamma$ decay and the appearance of the
tensor and the scalar form factors in the $K^+ \rightarrow \pi^o e^+ \nu$
decay.
\ \\

\section{Antisymmetric Tensor Fields}

The antisymmetric tensor fields have been discussed in literature for a long
time.$^{10}$ However, now they are used only in the supersymmetric theory of
gravitation.$^{11}$ Two types of second rank antisymmetric tensor fields exist.
The first ones play the role of the gauge fields and are described by a
gauge-invariant Lagrangian. The second ones are matter fields and their
Lagrangian is a conformal invariant one$^{12}$
\begin{equation}
{\cal{L}}_o = {1\over 4} \partial_{\lambda} T_{\mu\nu} \partial^{\lambda}
T^{\mu\nu} - \partial_{\mu} T^{\mu \lambda} \partial^{\nu} T_{\nu\lambda}.
\end{equation}
Henceforth we use these fields. We want to note here that this Lagrangian is
ghost free despite the widely spread belief that it is not. The six independent
components of the field $T_{\mu\nu}$ describe two massless particles: the
vector one $A_i = T_{oi}$ and the pseudovector one $B_i = {1\over 2} e_{ijk} T_{jk}$
$(i, j, k = 1, 2, 3)$. The particles must not be made massive by simple writing
a mass term $M^2T_{\mu\nu}T^{\mu\nu}$ by hand, because the Hamiltonian will not
be positively defined anymore. Therefore, the mechanism of the spontaneous
symmetry breaking must be used to make the fields massive. For this purpose it
is necessary to introduce an interaction of the tensor field with a massless
particle, that in its intermediate state has a pole $1/q^2$.$^{13}$ In contrast
to the usual Higgs mechanism of symmetry breaking in the case of a tensor
field it is necessary to use both a scalar field with a nonzero vacuum 
expectation value  $\sqrt 2\langle \varphi \rangle = M/2g$, 
and a massless vector particle
$V_{\mu}$, giving the pole $1/q^2$. Then the renormalizable gauge-invariant
interaction has the form:
\begin{equation}
{\cal{L}}_{SB} = 2g \varphi T_{\mu\nu} F^{\mu\nu} - 2g^2 \varphi^2
T_{\mu\nu} T^{\mu\nu},
\end{equation}
where $F_{\mu\nu} = \partial_{\mu} V_{\nu} - \partial_{\nu} V_{\mu}$. The
tensor field propagator, taking into account the diagrams in Figs.~1a and 1b, is:
\begin{equation}
\langle T(T_{\mu\nu}T_{\alpha\beta})\rangle_o
={2i\Pi_{\mu\nu\alpha\beta} \over {q^2 - M^2}},
\end{equation}
where
$$
\Pi_{\mu\nu\alpha\beta} (q) = {1\over 2} \left(
g_{\mu\alpha} g_{\nu\beta} - g_{\mu\beta} g_{\nu\alpha} \right) - {q_{\mu}
q_{\alpha} g_{\nu\beta} - q_{\mu} q_{\beta} g_{\nu\alpha} - q_{\nu} q_{\alpha}
g_{\mu\beta} + q_{\nu} q_{\beta} g_{\mu\alpha} \over {q^2}}.
$$
It is natural that the vector fields acquire mass by an exchange of a tensor
particle (Fig.~1c).

\setlength{\unitlength}{0.15cm}
\begin{center}
\begin{picture}(80,60)
\thicklines
\put(55,5){\Large Fig.~1.}
\put(18,28){\Large (a)}
\put(58,25){\Large (b)}
\put(18,-2){\Large (c)}
\put(0,0){\Large $V_{\mu}$}
\put(0,30){\Large $T_{\mu\nu}$}
\put(35,0){\Large $V_{\nu}$}
\put(35,30){\Large $T_{\alpha\beta}$}
\put(50,30){\Large $T_{\mu\nu}$}
\put(65,30){\Large $T_{\alpha\beta}$}
\put(7,20){\Large $\langle\varphi\rangle$}
\put(26,20){\Large $\langle\varphi\rangle$}
\put(7,50){\Large $\langle\varphi\rangle$}
\put(26,50){\Large $\langle\varphi\rangle$}
\put(47,50){\Large $\langle\varphi\rangle$}
\put(67,50){\Large $\langle\varphi\rangle$}
\multiput(10.5,5.5)(0,1){2}{\line(1,0){19}}
\multiput(0,35.5)(0,1){2}{\line(1,0){10.5}}
\multiput(29.5,35.5)(0,1){2}{\line(1,0){10.5}}
\multiput(10.5,6.5)(0,0.5){21}{\circle*{0.3}}
\multiput(29.5,6.5)(0,0.5){21}{\circle*{0.3}}
\multiput(10.5,15.5)(19,0){2}{\line(0,1){2}}
\multiput(9.5,16.5)(19,0){2}{\line(1,0){2}}
\multiput(10.5,6)(19,0){2}{\circle{1}}
\multiput(10.5,36.5)(0,0.5){21}{\circle*{0.3}}
\multiput(29.5,36.5)(0,0.5){21}{\circle*{0.3}}
\multiput(10.5,45.5)(19,0){2}{\line(0,1){2}}
\multiput(9.5,46.5)(19,0){2}{\line(1,0){2}}
\multiput(10.5,36)(19,0){2}{\circle{1}}
\multiput(1,6)(4,0){3}{\oval(2,2)[t]}
\multiput(31,6)(4,0){3}{\oval(2,2)[t]}
\multiput(12,36)(4,0){5}{\oval(2,2)[t]}
\multiput(3,6)(4,0){2}{\oval(2,2)[b]}
\multiput(33,6)(4,0){2}{\oval(2,2)[b]}
\multiput(14,36)(4,0){4}{\oval(2,2)[b]}
\multiput(50,35.5)(0,1){2}{\line(1,0){20}}
\put(60,36){\circle{1}}
\multiput(60,36.5)(0.5,0.5){21}{\circle*{0.3}}
\multiput(60,36.5)(-0.5,0.5){21}{\circle*{0.3}}
\multiput(50,45.5)(20,0){2}{\line(0,1){2}}
\multiput(49,46.5)(20,0){2}{\line(1,0){2}}
\end{picture}
\end{center}

The Lagrangian (3) is invariant under the axial transformation
$T^{\pm}_{\mu\nu} \rightarrow \exp(\pm i \theta) T^{\pm}_{\mu\nu}$,
where $T^{\pm}_{\mu\nu}$ are complex combinations of the real field
$T_{\mu\nu}$ and its dual field $\tilde {T}_{\mu\nu}={1\over 2}
e_{\mu\nu\alpha\beta}T^{\alpha\beta}$: $T^{\pm}_{\mu\nu} =
(T_{\mu\nu} \pm i \tilde {T}_{\mu\nu})/\sqrt 2$. The gauge fields can be
easily introduced by rewriting the Lagrangian (3) in terms of the charged
fields $T^{\pm}_{\mu\nu}$ and replacing the derivatives
$\partial^{\mu} T^{\pm}_{\mu\nu}$ by the covariant derivatives:
$D^{\mu} T^{\pm}_{\mu\nu} = (\partial^{\mu} \mp i A^{\mu}) T^{\pm}_{\mu\nu}$.
The Lagrangian has the form:
\begin{equation}
{\cal{L}} = - D^{\mu} T^+_{\mu\lambda} D^{\nu} T^-_{\nu\lambda}.
\end{equation}

In case where, besides the tensor and the axial gauge fields, a scalar and 
a vector gauge fields with an interaction of the type (4) are included,
the theory will then be plagued with chiral anomalies (Fig.~2). Usually,
in order to compensate these anomalies, additional fields with
opposite transformations  
$U^{\pm}_{\mu\nu} \rightarrow \exp(\mp i \theta) U^{\pm}_{\mu\nu}$
are introduced. This concerns the scalar fields as well (Fig.~2b).
Therefore, we can conclude that the extended electroweak model with
tensor fields can be constructed anomaly free if corresponding pairs
of tensor fields, like $T_{\mu\nu}$ and $U_{\mu\nu}$, are introduced.
The Higgs sector of (pseudo)scalar fields must be duplicated as well.

\setlength{\unitlength}{0.15cm}
\begin{center}
\begin{picture}(100,55)
\thicklines
\put(45,5){\Large Fig.~2.}
\put(20,14){\Large (a)}
\put(70,14){\Large (b)}
\put(0,39){\Large $A_{\lambda}$}
\put(42,53){\Large $V_{\mu}$}
\put(42,20){\Large $V_{\nu}$}
\put(50,39){\Large $A_{\lambda}$}
\put(92,53){\Large $V_{\mu}$}
\put(92,20){\Large $V_{\nu}$}
\multiput(8,36.5)(50,0){2}{\circle{1}}
\multiput(24,20.5)(50,0){2}{\circle{1}}
\multiput(24,52.5)(50,0){2}{\circle{1}}
\put(8,36){\line(1,-1){15.5}}
\put(8.5,36.5){\line(1,-1){15.5}}
\put(8,37){\line(1,1){15.5}}
\put(8.5,36.5){\line(1,1){15.5}}
\multiput(73.5,20.5)(1,0){2}{\line(0,1){32}}
\multiput(24,21)(0,0.5){63}{\circle*{0.3}}
\multiput(58,36)(0.5,-0.5){32}{\circle*{0.3}}
\multiput(58,37)(0.5,0.5){32}{\circle*{0.3}}
\multiput(0.5,36.5)(4,0){2}{\oval(2,2)[t]}
\multiput(2.5,36.5)(4,0){2}{\oval(2,2)[b]}
\multiput(50.5,36.5)(4,0){2}{\oval(2,2)[t]}
\multiput(52.5,36.5)(4,0){2}{\oval(2,2)[b]}
\multiput(25.5,52.5)(4,0){4}{\oval(2,2)[t]}
\multiput(27.5,52.5)(4,0){4}{\oval(2,2)[b]}
\multiput(75.5,52.5)(4,0){4}{\oval(2,2)[t]}
\multiput(77.5,52.5)(4,0){4}{\oval(2,2)[b]}
\multiput(25.5,20.5)(4,0){4}{\oval(2,2)[b]}
\multiput(27.5,20.5)(4,0){4}{\oval(2,2)[t]}
\multiput(75.5,20.5)(4,0){4}{\oval(2,2)[b]}
\multiput(77.5,20.5)(4,0){4}{\oval(2,2)[t]}
\end{picture}
\end{center}

\section{The Extended Electroweak Model}

We suppose a local $SU(2) \times U(1)$ electroweak symmetry with gauge
fields ${\bf{A}}_\mu$ and $B_{\mu}$. The fermion sector consists of
several generations (noted by the index $a$) two component spinors:
the lepton doublets $L_a=(\nu_L e_L)_a$ and the lepton singlets $e_{Ra}$;
the quark doublets $Q_a=(u_L'd_L')_a$ and the quark singlets $u_{Ra}'$,
$d_{Ra}'$. We \mbox{suppose that} neutrino is massless. Therefore, the mixing
of the generations concerns only quark sector. Let $u_{La}'=[S_u]_{ab}u_{Lb}$,
$u_{Ra}'=[T_u]_{ab}u_{Rb}$, $d_{La}'=[S_d]_{ab}d_{Lb}$ and 
$d_{Ra}'=[T_d]_{ab}d_{Rb}$, where the nonprimed fields are the eigenstates
of the mass matrix. Using the Kobayashi-Maskawa mixing matrix
$V_{ab}=[S^{\dag}_uS_d]_{ab}$ and introducing another two matrices
$U_{ab}=[S^{\dag}_uT_d]_{ab}$ and $W_{ab}=[T^{\dag}_uS_d]_{ab}$,
a mixing either for the up-type quarks or for the down-type quarks will result.

The tensor interaction does not conserve chirality, hence it couples the
left doublets and the right singlets of the fermion fields. In order to
have $SU(2) \times U(1)$ invariant Yukawa Lagrangian for the spinor and the
tensor fields, the tensor fields must be doublet ones. The requirement for
anomaly free extended model can be fulfilled by introducing two doublets of
scalar Higgs fields $H_1=(H^+_1 H^o_1)$ and $H_2=(H^o_2 H^-_2)$, and two
doublets of tensor fields $T_{\mu\nu}=(T^+_{\mu\nu} T^o_{\mu\nu})$
and $U_{\mu\nu}=(U^o_{\mu\nu} U^-_{\mu\nu})$ with opposite hypercharges:
$Y(T)=Y(H_1)=+1$, $Y(U)=Y(H_2)=-1$. Their interactions with the gauge fields
are introduced by the covariant derivative $D_{\mu} = \partial_{\mu} -
ig{\bf{T}}\cdot{\bf{A}}_\mu - ig'{Y\over 2}B_{\mu}$, where $g$ and $g'$
are the coupling constants, and ${\bf{T}}$ and $Y$ are the generators of the
$SU(2)$ and $U(1)$ groups, correspondingly. A triple interaction is also
allowed. It is of the type of the first term of the Lagrangian (4) and
it appears among the scalar fields, the tensor fields and the gauge fields
strength tensors $F_{\mu\nu}=\partial_{\mu}B_\nu - \partial_{\nu}B_\mu$,
${\bf{G}}_{\mu\nu}=\partial_{\mu}{\bf{A}}_\nu - \partial_{\nu}{\bf{A}}_\mu
+g{\bf{A}}_\mu\times{\bf{A}}_\nu$:
\begin{equation}
{\cal{L}}_3 = (f_1 H^{\dag}_1 T^{\mu\nu} + f_2 H^{\dag}_2 U^{\mu\nu})
\cdot F_{\mu\nu}
+(g_1 H^{\dag}_1 \mbox{\boldmath $\tau$} T^{\mu\nu}
+ g_2 H^{\dag}_2 \mbox{\boldmath $\tau$} U^{\mu\nu})
\cdot {\bf{G}}_{\mu\nu} + {\rm h.c.}
\end{equation}
The requirement for the photon to be massless after the symmetry breaking, when
\mbox{$\langle H^o_1\rangle=v_1$} and $\langle H^o_2\rangle=v_2$, 
sets the following relations $f_1/g_1=-f_2/g_2=\tan \theta_W$.

In the following we will discuss the tensor fields interaction to the fermions.
The $SU(2) \times U(1)$ invariant Lagrangian with Yukawa interactions is
\begin{equation}
{\cal{L}}_{Y} = {1\over 2} \left[ t^\ell_a\left(\bar{L}_a\sigma^{\mu\nu}
e_{Ra}\right) + t^q_a \left( \bar{Q}_a\sigma^{\mu\nu} d_{Ra}'\right) \right]
\left(\begin{array}{c} T^+_{\mu\nu} \\ T^o_{\mu\nu} \end{array} \right) +
{u^q_a\over 2} \left( \bar{Q}_a\sigma^{\mu\nu} u_{Ra}'\right)
\left(\begin{array}{cc} U^o_{\mu\nu} \\ U^-_{\mu\nu} \end{array} \right) + {\rm h.c.}
\end{equation}
We want to note here that if the right neutrino does not exist, the doublet
of the tensor field $U_{\mu\nu}$ interacts only with the quark fields.

After the spontaneuos symmetry breaking, mixing of the fields may result.
We are not going to discuss here the spontaneous symmetry breaking mechanism
in detail. We only need to note that the mass parameters appear from the 
interaction (7). In case of mixing, the propagator for the charged tensor
fields $T^\pm_{\mu\nu}$ and $U^\pm_{\mu\nu}$, will have the form:
\begin{equation}
{\cal{P}}(q)=\left( \begin{array}{cc}
\langle T(T^+T^-)\rangle_o & \langle T(T^+U^-)\rangle_o \\ 
\langle T(U^+T^-)\rangle_o & \langle T(U^+U^-)\rangle_o
\end{array} \right)=
{2i \over \Delta}\left( \begin{array}{cc}
(q^2-m^2)\Pi(q) & \mu^2 \\ \mu^2 & (q^2-M^2)\Pi(q)
\end{array} \right),
\end{equation}
where $\Delta=(q^2-m^2)(q^2-M^2)-\mu^4$.
Now we can write the effective Lagrangian for the lepton-quark interaction
due to the exchange of the charged tensor fields:
\begin{equation}
{\cal{L}}_{eff}=t^\ell_a\left(\bar{e}_{Ra}\sigma_{\mu\lambda}\nu_{La}\right)
{q^{\mu}q_{\nu} \over \Delta} \bar{u}_b \sigma^{\nu\lambda}\left[ 
(q^2-m^2)t^q_b (1+\gamma^5) U_{bc} - \mu^2 u^q_b  (1-\gamma^5) W_{bc}
\right]d_c +{\rm h.c.}
\end{equation}
The main requirement put on the interaction (10) is that it should not
contribute into the $\pi_{e_2}$ decay. Therefore, the pseudotensor terms 
$\bar{u}\sigma_{\mu\nu}\gamma^5d$, contributing to $\pi_{e_2}$ decay, must
cancel. Assuming for simplicity the coupling constant universality of the
tensor field interaction, i.e. $t^\ell=t^q=u^q$, and the mixing matrices $U$
and $W$ rather close to unity, in the static approximation $q^2 \ll \mu^2$,
$m^2$, $M^2$, there will be no contribution to the $\pi_{e_2}$ decay, provided
that $\mu^2=m^2$.

We shall discuss the interaction of the light quarks $u$, $d$ and $s$. In
order to avoid the flavor changing neutral currents we suppose
$S_{u,d}=T_{u,d}$. Then the mixing of the left and the right quarks 
$d_{L,R}'=\cos{\theta_C} d_{L,R} + \sin{\theta_C} s_{L,R}$,
$s_{L,R}'=- \sin{\theta_C} d_{L,R} + \cos{\theta_C} s_{L,R}$ is parametrized
by one Cabibbo angle $\theta_C$. The effective Lagrangian (10) for the
conserving strangeness case will have the form
\begin{equation}
{\cal{L}}_{\Delta S=0}
 = - {G_F \cos{\theta_C} \over {\sqrt 2}} f_t\; \bar{e} \sigma_{\mu\lambda}
(1-\gamma^5)\nu {q^{\mu}q_{\nu} \over q^2} \bar{u} \sigma^{\nu\lambda}d+{\rm h.c.}~,
\end{equation}
while that for the changing strangeness case will have the form
\begin{equation}
{\cal{L}}_{\Delta S=1}
 = - {G_F \sin{\theta_C} \over {\sqrt 2}} f_t\; \bar{e} \sigma_{\mu\lambda}
(1-\gamma^5)\nu {q^{\mu}q_{\nu} \over q^2} \bar{u} \sigma^{\nu\lambda}s+{\rm h.c.}~,
\end{equation}
where $f_t G_F /\sqrt 2 = t^2 / (M^2-m^2)$. The mass matrix in (9) will
have definite sign in case $m^2 \leq M^2$. This condition fixes the sign
of $f_t$: $f_t \geq 0$. As it will be shown below, this leads to the right
sign for the interference amplitude of the $\pi^- \rightarrow e^- \tilde{\nu}
\gamma$ decay.

\section{The $K^+ \rightarrow \pi^o e^+ \nu$ Decay}

The most general form of the matrix element for the
$K^+_{e_3}$ decay is$^{14}$
$$
M={G_F \sin{\theta_C} \over {\sqrt 2}}\; \bar{\nu} (1+\gamma^5)
\left\{ M_K f_S - {1 \over 2} \left[ ( P_K + P_{\pi})_{\mu} f_+
+ (P_K - P_{\pi})_{\mu} f_- \right] \gamma^{\mu} \right .
$$
\begin{equation}
\left . + i{f_T \over M_K} \sigma_{\mu\nu} P^{\mu}_K P^{\nu}_{\pi} \right\} e, 
\end{equation}
where the form factors $f_+$, $f_-$, $f_S$ and $f_T$ are functions only of 
the square of the four-momentum transfer to leptons $q^2=(P_K-P_{\pi})^2$.

In the relativistic quark model the form factor $f_T$ can be obtained from 
the diagram in Fig.~3. Substituting $ig\bar{u}\gamma^5 u \cdot \pi^0$ and
\mbox{$i\sqrt 2 g\bar{u}\gamma^5 s \cdot K^+$} into the quark-meson vertices
of the pion and the kaon correspondingly, and (12) into the quark-lepton vertex
we obtain
$$
M_T=iG_F \sin{\theta_C} f_t g^2 N_c\; \bar{\nu} \sigma_{\mu\lambda}(1+\gamma^5)e
{q^{\mu}q_{\nu} \over q^2}\mbox{\hspace{7cm}} 
$$
\begin{equation}
\mbox{\hspace{1cm}} \times \int {d^4l \over (2 \pi)^4} 
Sp \left[(\hat{l}-\hat{P_{\pi}}-M_u)^{-1} \gamma^5(\hat{l}-M_u)^{-1} \gamma^5
(\hat{l}-\hat{P_K}-M_s)^{-1} \sigma^{\nu\lambda} \right],
\end{equation}
where $M_u = 340$ MeV and $M_s = 490$ MeV are the constituent masses of $u$
and $s$ quarks. We want to note that the one-loop integral is a convergent
one and can be directly calculated. Finally, we obtain a relation between
the form factor $f_T$ and the constant $f_t$:
\begin{equation}
f_T = {\sqrt 2 \over 4 \pi^2} N_c {M_u a + (M_s - M_u) b \over M_K} g^2 f_t,
\end{equation}
where $N_c=3$ is the colour factor; $a$ and $b$ are defined by
$$
16\pi^2 M^2_K i \int {d^4l \over (2 \pi)^4} 
\left[(l-P_{\pi})^2-M^2_u \right]^{-1}\left[l^2-M^2_u \right]^{-1}
\left[(l-P_K)^2-M^2_s \right]^{-1} = a(q^2),
$$
\begin{equation}
16\pi^2 M^2_K i \int {d^4l \over (2 \pi)^4} l^{\mu} 
\left[(l-P_{\pi})^2-M^2_u \right]^{-1}\left[l^2-M^2_u \right]^{-1}
\left[(l-P_K)^2-M^2_s \right]^{-1} = 
\end{equation}
$$
= b(q^2)P^{\mu}_K + c(q^2)P^{\mu}_{\pi},
$$
and they depend weakly on $q^2$. Their average values are 
$a \approx 0.97$ and $b \approx 0.29$. 

With the help of the Goldberger-Treiman relation, the constant
$g$ can be expressed either through the pion decay constant $F_{\pi} = 131$ MeV
or the kaon decay constant
$F_K = 160$~MeV: $g \approx 2M_u/\sqrt 2 F_{\pi} \approx (M_u + M_s)/ 
\sqrt 2 F_K \approx 3.67$. Using the experimental data$^{15}$ $\mid {f_T/
f_+(0)} \mid = 0.38\pm0.11$ and taking $f_+(0) =\sqrt 2$ we obtain
$\mid f^K_t \mid = 0.49\pm0.14$.

\setlength{\unitlength}{0.15cm}
\begin{center}
\begin{picture}(50,45)
\thicklines
\put(20,0){\Large Fig.~3.}
\put(0,28){\Large $K^+$}
\put(42,39){\Large $\pi^o$}
\put(42,10){\Large $\nu$}
\put(42,18){\Large $e^+$}
\multiput(0,26)(0,1){2}{\line(1,0){8}}
\multiput(24,38)(0,1){2}{\line(1,0){16}}
\put(8,27){\vector(4,3){8}}
\put(24,14){\vector(-4,3){8}}
\put(24,14){\vector(4,-1){10}}
\put(24,38){\vector(0,-1){12.5}}
\put(40,18){\vector(-4,-1){8}}
\put(8,26){\line(4,-3){8}}
\put(24,14){\line(0,1){12.5}}
\put(24,14){\line(4,1){8}}
\put(24,39){\line(-4,-3){8}}
\put(40,10){\line(-4,1){6}}
\put(16,30){u}
\put(16,21){s}
\put(20,26){u}
\put(12,33){\bf $l$}
\put(25,26){\bf $l-P_\pi$}
\put(10,17){\bf $l-P_K$}
\end{picture}
\end{center}

The tensor interaction (12) cannot explain the value$^{15}$  $\mid {f_S/f_+} \mid
=0.084\pm0.023$  and the relative phase angle value$^2$  $\phi_{st}
=0.00\pm0.52$  of the scalar form factor in (13). The account of the
radiative corrections to (12) leads to an insufficient value for 
$\mid {f_S/f_+} \mid$ :$\mid {f_S/f_+} \mid \sim 10^{-3}$ and $\phi_{st}=\pi$.
Hence in our model two Higgs doublets exist, an interaction due to
charged Higgs fields is possible:
\begin{equation}
{\cal{L}}_S=
- {G_F \sin{\theta_C} \over {\sqrt 2}} f_s\; \bar{e} (1-\gamma^5)\nu
\cdot \bar{u} s + {\rm h.c.}
\end{equation}
Here as well, in order to avoid the contribution to the $\pi_{e_2}$ decay,
the quark pseudoscalar term must be absent. Then
$$
M_S=iG_F \sin{\theta_C} f_s g^2 N_c\; \bar{\nu} (1+\gamma^5)e\mbox{\hspace{7cm}} 
$$
\begin{equation}
\mbox{\hspace{1cm}} \times \int {d^4l \over (2 \pi)^4} 
Sp \left[(\hat{l}-\hat{P_{\pi}}-M_u)^{-1} \gamma^5(\hat{l}-M_u)^{-1} \gamma^5
(\hat{l}-\hat{P_K}-M_s)^{-1} \right].
\end{equation}
The leading diverging part of the integral in (18) has been evaluated using
the quark loop for the kaon decay $K \rightarrow e \nu$ 
\begin{equation}
2(M_u+M_s)\sqrt 2 gN_c \int {d^4l \over (2 \pi)^4} 
\left[l^2-M_u^2 \right]^{-1} \left[(l-p)^2-M_s^2 \right]^{-1} = iF_K.
\end{equation}
The following relation is obtained: $\mid f_S/f_+ \mid=f_s M_s/M_K$.
Then the coupling constant of the scalar interaction (17) is 
$\mid f_s\mid=0.086\pm0.023$.

\section{The $\pi^- \rightarrow e^- \tilde{\nu} \gamma$ Decay}

  We shall use the same parametrization of the matrix element as the one in
Ref.~16
\begin{equation}
M = M_{IB} + M_{SD},
\end{equation}
where
\begin{equation}
M_{IB}=-i{e G_F \cos\theta_C \over \sqrt 2} F_{\pi} m_e \epsilon_{\mu}\; 
\bar{e} \left[ \left({k \over kq} - {p \over pq}\right)^{\mu} 
- {i \sigma^{\mu\nu}q_{\nu}
\over 2kq} \right] (1-\gamma^5) \nu_e
\end{equation}
is a QED correction to the $\pi \rightarrow e \nu$ decay (inner bremsstrahlung)
and
\begin{equation}
M_{SD}=-{e G_F \cos\theta_C \over \sqrt 2 M_{\pi}} \epsilon^{\mu}
\left[ F_V e_{\mu\nu\rho\sigma}p^{\rho}q^{\sigma} - iF_A(pq \cdot g_{\mu\nu}
-p_{\mu}q_{\nu}) \right] \bar{e} \gamma^{\nu}(1-\gamma^5)\nu_e
\end{equation}
is a structure-dependent amplitude parametrized by two form factors $F_V$
and $F_A$; $\epsilon^{\mu}$ is the photon polarization vector; $p$, $k$ and $q$
are the pion, electron and photon four-momenta respectively.

\setlength{\unitlength}{0.15cm}
\begin{center}
\begin{picture}(100,50)
\thicklines
\put(47,0){\Large Fig.~4.}
\put(0,28){\Large $\pi^-$}
\put(42,39){\Large $\gamma$}
\put(42,10){\Large $e^-$}
\put(42,18){\Large $\tilde{\nu}$}
\put(50,28){\Large $\pi^-$}
\put(92,14){\Large $\gamma$}
\put(92,35){\Large $e^-$}
\put(92,43){\Large $\tilde{\nu}$}
\multiput(0,26)(0,1){2}{\line(1,0){8}}
\put(8,27){\vector(4,3){8}}
\put(24,14){\vector(-4,3){8}}
\put(24,14){\vector(4,-1){10}}
\put(24,39){\vector(0,-1){12.5}}
\put(40,18){\vector(-4,-1){8}}
\put(8,26){\line(4,-3){8}}
\put(24,14){\line(0,1){12.5}}
\put(24,14){\line(4,1){8}}
\put(24,39){\line(-4,-3){8}}
\put(40,10){\line(-4,1){6}}
\multiput(25,39)(4,0){4}{\oval(2,2)[t]}
\multiput(27,39)(4,0){4}{\oval(2,2)[b]}
\multiput(50,26)(0,1){2}{\line(1,0){8}}
\put(58,27){\vector(4,3){8}}
\put(74,14){\vector(-4,3){8}}
\put(74,39){\vector(4,-1){10}}
\put(74,39){\vector(0,-1){12.5}}
\put(90,43){\vector(-4,-1){8}}
\put(58,26){\line(4,-3){8}}
\put(74,14){\line(0,1){12.5}}
\put(74,39){\line(4,1){8}}
\put(74,39){\line(-4,-3){8}}
\put(90,35){\line(-4,1){6}}
\multiput(75,14)(4,0){4}{\oval(2,2)[b]}
\multiput(77,14)(4,0){4}{\oval(2,2)[t]}
\put(16,30){d}
\put(16,21){u}
\put(20,26){d}
\put(66,30){d}
\put(66,21){u}
\put(70,26){u}
\put(12,33){\bf $l$}
\put(25,26){\bf $l-q$}
\put(10,17){\bf $l-p$}
\put(58,33){\bf $l+p$}
\put(75,26){\bf $l+q$}
\put(62,17){\bf $l$}
\end{picture}
\end{center}

\ \\

Using the new interaction (11), the additional to (20) matrix element can be
calculated in the framework of the relativistic quark model (Fig.~4)
$$
M_T= eG_F \cos{\theta_C} f_t gN_c(e_d+e_u)\epsilon^{\mu}\;
\bar{e} \sigma^{\gamma
\beta}(1-\gamma^5)\nu_e {(p-q)_{\gamma}(p-q)^{\alpha} \over (p-q)^2}
\mbox{\hspace{2cm}}
$$
\begin{equation}
\times \int {d^4l \over (2 \pi)^4} 
Sp \left[(\hat{l}-\hat{q}-m)^{-1} \gamma_{\mu}(\hat{l}-m)^{-1} \gamma^5
(\hat{l}-\hat{p}-m)^{-1} \sigma_{\alpha\beta} \right]
\end{equation}
here $m \equiv M_u \approx M_d$ is the constituent mass of $u$ and $d$ quarks.
The leading diverging part of the integral in (23) has been evaluated using
the quark loop for the pion decay $\pi \rightarrow e \nu$ 
\begin{equation}
4m\sqrt 2 gN_c \int {d^4l \over (2 \pi)^4} 
\left[l^2-m^2 \right]^{-1} \left[(l-p)^2-m^2 \right]^{-1} = iF_{\pi}.
\end{equation}
Therefore, the additional amplitude reads
\begin{equation}
M_T={eG_F \cos{\theta_C} \over \sqrt 2} F_T
{1 \over 2} \left[\epsilon^{\mu}q^{\nu} +
{(\epsilon p)q^{\mu} - (pq)\epsilon^{\mu} \over {(p-q)^2}}(p-q)^{\nu}\right]
\bar{e}\sigma_{\mu\nu} (1-\gamma^5)\nu_e.
\end{equation}
The constant $F_T$ is expressed by $f_t$ in a similar way as in Ref.~4:
\begin{equation}
F_T=(e_d+e_u){F_{\pi} \over m}f_t.
\end{equation}

Neglecting the electron mass, the decay amplitude with the leading interference
term has the form:
$$
{d^2\Gamma \over dx d\lambda}={\alpha \over 2 \pi}\Gamma_{\pi \rightarrow e\nu}
\left\{ IB(x,\lambda) + a^2_{SD} \left[ (F_V+F_A)^2 SD^+(x,\lambda)+
(F_V-F_A)^2 SD^-(x,\lambda) \right]	\right .
$$
\begin{equation}
 - a_{SD} F_T I(x,\lambda) \Big\}
\end{equation}
where $a_{SD}=M^2_{\pi}/2F_{\pi}m_e$,
\begin{eqnarray*}
IB(x,\lambda)={1-\lambda \over \lambda}{(1-x)^2+1 \over x},
\mbox{\hspace{1cm}}
SD^+(x,\lambda)=\lambda^2(1-x)x^3,	
\\
SD^-(x,\lambda)=(1-\lambda)^2(1-x)x^3,\mbox{\hspace{2.2cm}}
I(x,\lambda)=(1-\lambda)x^2.
\end{eqnarray*}
In the pion frame the variables $x$ and $\lambda$ are $x=2E_{\gamma}/
M_{\pi}$, $\lambda=2E_e/ M_{\pi}\sin^2{\theta_{e\gamma} \over 2}$.

The theoretical branching ratio, calculated within the standard $V\!-\!A$ model
for the kinematical region $0.3<x<1.0$, $0.2<\lambda<1.0$ is $B^{th}=
(2.41\pm0.07)\times 10^{-7}$. The discrepancy between the experimental total
decay probability $B^{exp}=(1.61\pm0.23)\times 10^{-7}$ and the calculated
one $B^{th}$ is due to the negative value of $SD^-$. As far as the
distributions of the spectra for $I(x,\lambda)$ and $SD^-(x,\lambda)$ terms
are approximately similar ones,
this discrepancy can be avoided if $F_T=(3.72\pm1.20)\times 10^{-2}$.
From (26) we can obtain $f^{\pi}_t=0.29\pm0.09$.

\section{Conclusions}

From the calculation of the constants $f^{\pi}_t$ and $f^K_t$, on the
basis of the meson decay experiments, it shows that they may be equal. Therefore,
we can conclude that a universal effective interaction may exist
\begin{equation}
{\cal{L}}_T
 = - {G_F \over {\sqrt 2}} f_t\; \bar{e} \sigma_{\mu\lambda} (1-\gamma^5)\nu
 {q^{\mu}q_{\nu} \over q^2} \bar{u} \sigma^{\nu\lambda}d' + {\rm h.c.}
\end{equation}
with an average  coupling constant $f_t=0.39\pm0.18$, where
$d'=\cos{\theta_C}d+\sin{\theta_C}s$.
The absence of the quark pseudotensor terms in (28) allows to avoid 
the possible anomalously great contribution of the radiative
corrections to the $\pi_{e_2}$ decay.
The effective scalar interaction
\begin{equation}
{\cal{L}}_S
 = - {G_F \over {\sqrt 2}} f_s\; \bar{e} (1-\gamma^5)\nu\cdot\bar{u} d' + {\rm h.c.}
\end{equation}
does not contribute to the $\pi_{e_2}$ decay as well. As far as (29) gives no
contribution in $\pi^- \rightarrow e^- \tilde{\nu} \gamma$ decay, the value of
the scalar constant $\mid f_s\mid=0.086\pm0.023$ is obtained only from the
$K^+ \rightarrow \pi^o e^+ \nu$ decay. Unfortunately, we cannot say anything
about the values of the Yukawa coupling constants of the tensor fields
without using higher symmetries. Therefore, we will abstain from conclusions
concerning the tensor particle masses.

\section*{Acknowledgments}

I am grateful to L.~Avdeev, L.~Litov and Professor M.~Mateev for useful
discussions and help. This work is financially supported by Grant-in-Aid
for Scientific Research F-214/2096 from the Bulgarian Ministry of Education,
Science and Culture.

\section*{References}

\begin{description}
\item[ 1.] V.~N.~Bolotov et al., \underline{Phys. Lett.} B243 (1990) 308.
\item[ 2.] S.~A.~Akimenko et al., \underline{Phys. Lett.} B259 (1991) 225.
\item[ 3.] H.~Stainer et al., \underline{Phys. Lett.} B36 (1971) 521.
\item[ 4.] A.~A.~Poblaguev, \underline{Phys. Lett.} B238 (1990) 108.
\item[ 5.] V.~M.~Belyaev and Ian I.~Kogan, \underline{Phys. Lett.} B280 (1992) 238.
\item[ 6.] A.~A.~Poblaguev, \underline{Phys. Lett.} B286 (1992) 169.
\item[ 7.] M.~B.~Voloshin, \underline{Phys. Lett.} B283 (1992) 120.
\item[ 8.] P.~Depommier et al., \underline{Phys. Lett.} 7 (1963) 285; \\
    A.~Stetz et al., \underline{Nucl. Phys.} B138 (1978) 285; \\
    A.~Bay et al., \underline{Phys. Lett.} B174 (1986) 445; \\
    L.~E.~Piilonen et al., \underline{Phys. Rev.} Lett. 57 (1986) 1402; \\
    S.~Egly et al., \underline{Phys. Lett.} B175 (1986) 97; B222 (1989) 533. 
\item[ 9.] B.~A.~Campbell and K.~A.~Peterson, 
    \underline{Phys. Lett.} B192 (1987) 401; \\
    O.~Shankar, \underline{Nucl. Phys.} B204 (1982) 375.    
\item[10.] N.~Kemmer, \underline{Helv. Acta}, 33 (1960) 829; \\
    V.~I.~Ogievetsky and I.~V.~Polubarinov, \underline{Yadr. Fiz.} 4 (1966) 216.
\item[11.] \underline{Supergravities in Diverse Dimensions}, eds. A.~Salam and
    E.~Sezgin \\(North-Holland and World Scientific, 1989).
\item[12.] B. de Wit and J.~W. van Holten, \underline{Nucl. Phys.} B155 (1979) 530.
\item[13.] J.~Schwinger, \underline{Phys. Rev.} 125 (1962) 397; \\
    F.~Englert and R.~Brout, \underline{Phys. Rev. Lett.} 13 (1964) 321.
\item[14.] H.~Braun et al., \underline{Nucl. Phys.} B89 (1975) 210.
\item[15.] \underline{Particle Data Group, Phys. Rev.} D45, Part 2 (June 1992).
\item[16.] D.~A.~Bryman, P.~Depommier and C.~Leroy, 
    \underline{Phys. Rep.} 88 (1982) 151.
\end{description}

\end{document}